\documentclass[preprint,aps,amsfonts, amssymb, amsmath, prb]{revtex4-2}
\usepackage{bm}
\usepackage{graphicx}

\begin{document}

\title{Nonlocal thermal noise in electrically coupled conductors: A microscopic two-dimensional study}

\author{Jorge Berger}
\affiliation{Department of Physics, Braude College, Karmiel, Israel
}


\begin{abstract}
The Johnson-Nyquist theory is commonly implemented by representing a conductor as a collection of independent local thermal-noise sources whose strength is determined by the local temperature. Whether this local-noise representation remains valid for electrically coupled conductors maintained at different temperatures has received comparatively little attention. We investigate this question by means of microscopic two-dimensional simulations of interacting charge carriers in conducting wires capacitively coupled. The model reproduces Ohm's law, the equilibrium Johnson noise, and vanishing correlations between detached wire segments when both wires are at the same temperature. However, when the wires are held at different temperatures,  finite correlations develop between the electromotive forces generated in distant segments, leading to systematic deviations of the Johnson temperature inferred from the local-noise picture. The effect persists although the microscopic particle interactions are short-ranged and the two wires interact only through the capacitive coupling. These results suggest that the independent-local-source representation of thermal noise may not remain valid in electrically coupled nonequilibrium conductors.
\end{abstract}

\maketitle

\section{Introduction}
Can the thermal noise generated in one conductor depend on the temperature of another conductor to which it is coupled only capacitively? Electric noise, due to thermal motion of charged particles, was expected from kinetic theory \cite{ein}. It was measured by Johnson \cite{john} in 1928 and explained by Nyquist \cite{ny}. 

Following Nyquist, a conductor with resistance $R$ at temperature $T$ behaves as a power supply with internal resistance $R$ that generates an emf (electromotive force) $\tilde{\epsilon}$. Expressed in the time domain, if $\epsilon$ is the time average of $\tilde{\epsilon}$ over a duration $\Delta t$ and $\Delta t\gg \hbar/k_BT$, where $k_B$ is the Boltzmann constant, then the variance of $\epsilon$ (over an ensemble of measurements with duration $\Delta t$) is
\begin{equation}
    \langle \epsilon^2\rangle =2k_BTR/\Delta t
 \label{varep}\,.
\end{equation}
Since the only property of the conductor involved in this variance is its resistance, Eq.\ (\ref{varep}) provides an ideal thermometric procedure \cite{Qu}.

Nyquist's result was obtained from thermodynamic arguments and therefore a generalization is required for the description of thermal noise if the temperature is not uniform. From Nyquist's original wording it may be understood that the emf generated by a conductor only depends on the agitation of the charges \textit{inside} the conductor. We dub this view ``the Nyquist scenario". If the emf generated by a conductor is independent of the temperature beyond its boundaries, then the conductor can be divided into small regions, each with a sufficiently uniform temperature $T(\vec{r})$ and can be regarded as a small conductor. It follows that each region generates an emf with variance given by Eq.\ (\ref{varep}), the emf's are uncorrelated, and the variance of the resultant emf is given by their sum if the regions are positioned in series.  

Temperature nonuniformity may appear not only as a function of position but also as a difference between components. For example, if electrons collide among themselves much more frequently than with phonons, their temperature may be significantly different from that of the phonons \cite{Urbina}. In this case, the relevant temperature in Eq.\ (\ref{varep}) is that of the charged particles, i.e.\ the electrons. Temperature differences have also been found between spin-up and spin-down electrons \cite{Dej}.

Sukhorukov and Loss (SL) \cite{Loss} studied a multi-temperature situation: a diffusive conductor with a set of terminals such that at terminal $i$ electrons enter or exit the conductor with local temperature $T_i$, whereas the rest of the conductor boundary is impermeable to electrons and heat. Inside the conductor, electrons undergo elastic impurity scattering, so the lattice temperature is irrelevant. SL note that inelastic scattering can be added to the collision integral, but in the diffusive limit considered this does not affect momentum relaxation and hence the current density. They define a local effective electronic noise temperature inside the conductor, which depends on geometry and on the temperatures and voltages at the terminals. The SL approach is typically used in the analysis of multiprobe noise measurements in conductors with low electron-phonon coupling, such as graphene \cite{Brian}. Since the SL approach relates noise strength to heat transport, it is also invoked in junctions where the temperature changes abruptly as a function of position \cite{Nature}.

From a modern perspective, representation of thermal noise by independent local Langevin sources is justified in equilibrium situations by the fluctuation-dissipation theorem (FDT) \cite{Kubo}. For any considered system, FDT relates every fluctuation to the agents that generate and dissipate it. However, in nonequilibrium steady states the validity of this locality assumption is less clear. 
 A noteworthy case is provided by hydrodynamic systems, in which non-equilibrium fluctuations dramatically larger than in equilibrium have been observed \cite{Jan}. In the hydrodynamic case, the justification for the extension of FDT to nonequilibrium is addressed in Chapter 3 of \cite{hydro}. 
The present work considers a different situation: the temperatures of two electrically coupled conductors are externally fixed. The question addressed here is whether electrical coupling and short-range particle-particle interaction alone can give rise to correlations that invalidate the local-noise description.

In a previous study \cite{BK} we advanced the idea that due to the repulsion between electrons, sufficiently strong to establish almost perfect electroneutrality over long distances, the emf generated in a conductor depends also on the temperature of other resistors electrically connected to it. We considered the opposite limit to that of SL, in which the electronic energy distribution is reset to the temperature of the local bath after every collision. Experiments that correspond to this limit are available \cite{Bellon,prl,Berut}, but were not tailored to observe our prediction. We therefore performed simulations, which indeed support our prediction.

The main limitation of \cite{BK} is that the simulations in it are strictly 1D. In this case, the probability of a particle passing another is negligible, giving rise to a correlation that is not necessarily present in a thin wire. Another inconvenience of \cite{BK} is its focus on closed circuits, in which the measurement of enf is not straightforward; here, we will focus on measurable voltages.
The present study includes 2D simulations that aim to investigate thermal noise in circuits of resistors at different temperatures connected by ideal capacitors.

\section{Model}
\subsection{Choice of the model\label{choice}}
We consider a system of $N$ classical particles with mass $m$ and charge $q$, intended to resemble the behavior of electrons in a wire made of metal or an extrinsic semiconductor. The particles should therefore repel each other (isotropically in the simplest case) and this repulsion should be screened for long distances. We also require a computationally friendly interparticle interaction: it should vanish for most particle pairs, be continuous with continuous space derivatives, and expressible by means of computationally ``cheap" operations. Denoting by $\vec{r}$ the position of particle $j$ relative to particle $i$, all these requirements are satisfied if the force exerted by $i$ on $j$ is
\begin{equation}
    \vec{F}=\begin{cases}
    (3k_c\ell^4/8)(\ell/r-1/2)\vec{r} & r\le \ell/2\\  
    k_c(\ell^2-r^2)^2\vec{r} & \ell/2<r\le \ell\\
    \vec{0} & r>\ell
    \end{cases}
    \label{myforce}
\end{equation}
where $\ell$ and $k_c$ are parameters.

The particles also interact with lattice deformations (phonons and defects), and every time $\tau$ each particle collides with a deformation (all particles simultaneously). Immediately after a collision, every particle acquires Maxwellian distribution, i.e.\ each component of its velocity has a random value, with zero average, variance $k_BT/m$, and normal distribution.

Lengths, times, voltages, currents, temperatures, and $k_c$ will, respectively, be taken in units of $\ell$, $\tau$, $m\ell^2/q\tau^2$, $q/\tau$, $m\ell^2/k_B\tau^2$ and $m/\ell^4\tau^2$; in cases where writing these units explicitly would be cumbersome, they will be omitted.

We mimic a ``wire" by imposing that the position $(x_i,y_i)$ of any particle $i$ be in the rectangle $0\le x_i\le L$, $0\le y_i\le W$, with $L\gg W$. The four walls repel the particles near them. We take the repulsion forces as if there were a mirror charged particle at the other side of the wall (that interacts only with the mirrored particle).

If a particle hits a lateral wall at $y=0,W$, we impose an elastic collision (the longitudinal component of the velocity, $v_x$, is unchanged and $v_y$ changes sign). However, in order to soften the influence of finite length, when a particle hits an end at $x=0,L$, we assume instead that it is absorbed and immediately re-emitted. The re-emission velocity distribution is dictated by the temperature $T_w$ of the wall at that end: the new longitudinal component of the velocity is $v_x=\pm\sqrt{-(2k_BT_w/m)\log X}$, where $X$ is a random variable uniformly distributed between 0 and 1 and the sign of $v_x$ directs into the interior of the rectangle, while $v_y$ has Maxwellian distribution. For $T_w$ equal to the temperature within the rectangle it confines, and provided that the velocity change during the last step in which the particle reaches the wall can be neglected, this distribution obeys detailed balance with a Maxwellian distribution within the rectangle. After stabilization, 
we used a simpler method that guarantees detailed balance: the re-emission velocity was $(v_x,v_y)=(-v'_x,v'_y)$, where $v'_x$ ($v'_y$) is the $x$- ($y$-) velocity component of the impinging particle that hit the wall 100 (99) collisions ago.

Our model does not involve generation of electromagnetic field by the motion of a charge. This is a low-frequency assumption, expected to be valid for $\Delta t\gg L/c$, where $c$ is the speed of light.

\subsection{Rigidity and electroneutrality}
We still have to choose the parameters of the model. Some of these parameters have been taken as units. The width $W$ should be as small as possible while leading to multi-dimensional behavior; as will be discussed below, $W=2.2\ell$ is sufficient for this purpose. The remaining parameters may be lumped into an effective interparticle strength, $Nk_c\ell^7/Lk_BT$.

Since we want to mimic electrons in a solid, we would like our system to behave as an incompressible fluid that obeys Kirchhoff's law. By ``Kirchhoff's law" we mean the uncorrected law that refers to flow of particles only and does not invoke ``displacement current".
Electrons withstand compression because they obey the Pauli principle, 
because their repulsion force diverges as $r\to 0$, and because they interact with a large number of particles. We are therefore interested in a large interaction strength between the particles. On the other hand, a large density of particles demands evaluation of many forces per particle, and large $k_c$ requires small evolution steps.

\begin{figure}[tbh]
\scalebox{0.85}{\includegraphics{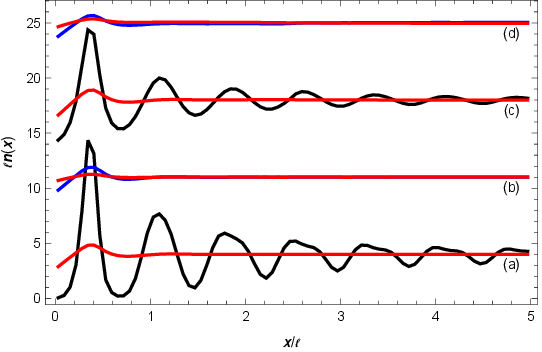}}
\caption{\label{oscillations}Time-average of the density of particles in a third of a wire close to one of its ends. Black: $T=0.25$; blue: $T=1$; red: $T=3.5$. (a) $k_c=50$; (b) $k_c=15$; (c) like (a), but the hot and the cold wire are connected by capacitors, as in Fig.\ref{configurations}(a); (d) like (b), with the wires connected as in Fig.\ref{configurations}(a). In all cases $N=60$ and $L=15\ell$. 
For clarity,  lines (b)–(d) are vertically shifted by 7, 14, and 21, respectively. }
\end{figure}

An additional complication of large interparticle strength is the crossover to a liquid‑like phase \cite{liquid} with short‑range order (that is not homogeneous). In Fig.\ \ref{oscillations}(a) we see that, for low temperature and large interparticle interaction, the density $n(x)$ of particles (per length) has strong oscillations near the ends of the wire, with maxima that appear quasiperiodically at distances of $\sim 0.8\ell$. Figure \ref{oscillations}(a) shows that these spatial oscillations are smaller for higher temperatures, and comparison with Fig.\ \ref{oscillations}(b) shows that oscillations are smaller for smaller $k_c$. In order to be on the safe side, we will generally take $N/L=4/\ell$, $k_c=15$, and temperatures in the range $1\le T\le 3.5$.

\subsection{Definitions}
We assume that experiments will not measure $y$-dependencies and will keep track of quantities ``along" the wire only. For an incompressible charged fluid, the current is independent of position. Our model does not guarantee incompressibility and we therefore define the instantaneous current $\tilde{I}$ by means of a spatial average of the current density:
\begin{equation}
\tilde{I}:=(q/L)\sum_{i=1}^N v_{i,x}\;.
\label{defI}
\end{equation}
The measured current $I$ is obtained by averaging $\tilde{I}$ over the measurement duration. We note that (imagining the resistor as straight) the current equals the total longitudinal momentum of the particles multiplied by $q/mL$.

Voltage is the agent that pushes the charged particles to maintain a current despite the randomization action of collisions. Following the reasoning presented in \cite{BK} and denoting the instantaneous voltage by $\tilde{V}$, $\tilde{V}(t)dt$ is the change in the momentum of the particles during the short period of time $dt$, multiplied by $(L/qN)$, i.e.\
\begin{equation}
    \tilde{V}(t)dt:=\frac{L}{qN}\left[\sum F_x^{\rm ext}dt+m\left(\sum v_x^{\rm in}-\sum v_x^{\rm out}  \right)\right]\,.
    \label{defV}
\end{equation}

Equations (\ref{defI})-(\ref{defV}) [as well as  Eq.\ (\ref{Drude})  below] may refer to the whole wire, but may also refer to a segment of the wire and, in that case, $L$ and $N$ should be replaced by the corresponding length and number of particles in the segment. $\sum F_x^{\rm ext}$ is the total longitudinal force (excluding collisions against lattice distortions) exerted on the particles; for the entire wire it is exerted by the walls at $x=0,L$, and in the case of a segment it includes the forces exerted by the particles in the adjacent segments. $\sum v_x^{\rm in}$ ($\sum v_x^{\rm out}$) is a sum over the particles that enter (leave) the considered segment during the lapse of time $dt$; for a particle that rebounds at $x=0,L$, the impinging (ejected back) particle may be regarded as a leaving (entering) particle. 

The measured voltage $V$ is a weighted average of $\tilde{V}(t)$ over the duration of the measurement. The reason for the weight is the ``loss of memory" of the current after a collision. Therefore, denoting by $t_r$ the remaining time until the next collision with the lattice deformations, the weight should decrease as $t_r$ decreases. In addition to collisions of all the particles every time $\tau$, there are particles that collide against the walls at $x=0,L$. Since for the parameters that we have considered the number of these collisions is smaller by more than an order of magnitude than that of collisions against deformations, and since we have not found correlation between $\tilde{V}(t)$ and $t_r$, we take the weight that is appropriate for collisions against deformations only, $2t_r/\tau$. 

The emf is defined as
\begin{equation}
    \epsilon =V+RI \,,
    \label{defep}
\end{equation}
where $R$ is the resistance of the wire or segment considered. For a Drude model in which collisions are a periodic event,
\begin{equation}
 R=2mL^2/q^2N\tau\;.
 \label{Drude}
 \end{equation}
 
 In the case of a segment, $N$ fluctuates and hence $R$ fluctuates. If  $Nk_c\ell^6\gg k_BT$, fluctuations are small \cite{BK} and $N$ can be replaced by its average value. However, if we suspect that the density of particles may not be uniform, (\ref{Drude}) has to be replaced with
\begin{equation}
 R=\frac{2m}{q^2\tau}\int_{x_1}^{x_2}\frac{dx}{n(x)}\;,
 \label{nonDrude}
 \end{equation}
 where $x_{1,2}$ are the ends of the considered segment. In fact, even if the number of particles between $x_1$ and $x_2$ is fixed, $n(x)$ and hence $R$ can fluctuate.

\subsection{Tests of the model}
Before using this model to investigate circuits with resistors at different temperatures, we verified that it reproduces the known results in the standard cases.

\subsubsection{Ohm's law \label{Ohm}}
We want to test Ohm's law and Drude's model on a finite wire of length $L$. dc current cannot flow in an open circuit but, since electroneutrality is not perfect, ac currents are possible. We therefore invoke an external agent that applies a periodic force $A\cos\omega t$ on each particle in the range $0.25L<x<0.75L$ (we prefer to keep the regions near the ends untouched). As a consequence of these forces, ac voltages and currents build up. We can then use (\ref{defI}) and (\ref{defV}) to evaluate the current and the voltage as functions of time and obtain their $\omega$-components. Dividing the voltage phasor by the current phasor, we finally obtain the impedance $Z(\omega)$. In this test, we neglected in Eq.\  (\ref{defV}) the fluctuations in the number of particles in $(0.25L,0.75L)$, and took it as its average during $100\tau$. 

\begin{figure}[tbh]
\scalebox{0.85}{\includegraphics{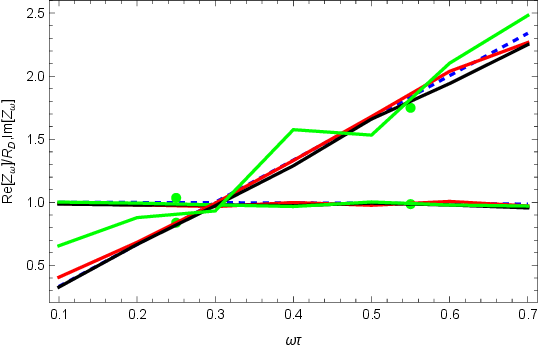}}
\caption{\label{ImZw} Impedance of a wire as a function of $\omega\tau$. Red: $T=3.5$; black: $T=0.25$; blue (dashed): values predicted by (\ref{Y}); green lines: impedance when a capacitor is inserted at $x=L/2$, for $T=1$. To keep scales similar, only Re$[Z_\omega]$ was divided by the Drude resistance,$R_D$. Parameters: $N=80$, $L=20$, $W=2.2$, $k_c=15$; $C'=1/k_c$ (if capacitor is present). Green dots: Impedance for $N=160$, $L=40$ (for this comparison, Im[$Z$] was divided by 2). }
\end{figure}

Figure \ref{ImZw} shows the impedances that we obtained for a particular wire, for $T=0.25$ and for $T=3.5$. These impedances are compared with the values predicted by Eq.\ (\ref{Y}). The amplitude $A$ of the force applied by the external agent was of the order of 0.2. The amplitudes of the voltages that built up for $T=3.5$ are roughly twice those that built up for $T=0.25$ and, when comparing $\omega\tau=0.1$ with $\omega\tau=0.7$, the ratio is of an order of magnitude; however, the resistance is practically the same in all cases. A representative sampling indicates that the largest source of deviation of our results from (\ref{Y}) was the size of the evolution step, which in this test was $0.01\tau$.

\subsubsection{Multidimensional behavior}
In our previous study \cite{BK} we considered mathematically 1D wires. In that case particles are practically unable to pass by another, leading to the suspicion that the correlations found are an artifact of this inability, analogous to the case of a Luttinger liquid \cite{Giamarchi}. To remove this suspicion, we want to make sure that, during a measurement, every particle has a significant probability of passing by another particle (along the longitudinal direction). Explicitly, if $\Delta t$ is the measurement duration and $\tau_P$ is the average time required for a particle to pass by another particle, our requirement is $\tau_P\alt\Delta t$.

Let us estimate $\tau_P$ for a particular set of parameters. For $k_c=15$ and $T=0.25$, a typical particle has kinetic energy $k_BT/2=0.125$ (for motion in either $x$ or $y$-direction and relative to the lattice). Using (\ref{myforce}), we can numerically estimate that the center of this particle can approach the center of another particle (or its reflection on a wall) within the range beyond a distance of $\sim 0.8\ell$ and, accordingly, the probability of passing is roughly that of rigid balls with diameter $0.8\ell$. For a width $W=2.2\ell$, the centers of such rigid balls are constrained within the region $0.4\ell\le y\le 1.8\ell$ and, denoting by $y$ the lateral position of one of the particles, the probability of passing in a single attempt is $(1/1.4^2)(\int_{0.4\ell}^\ell (\ell-y)dy+\int_{1.2\ell}^{1.8\ell} (y-1.2\ell)dy)\approx 0.2$. The probability of having passed after $n$ attempts is therefore $P(n)\approx 1-0.8^n$. For a density $N/L=4/\ell$ there are roughly two attempts during a time $\tau$, so the passing probability during $\tau$ is roughly 0.4 and during, say, $4\tau$, the probability becomes 0.8.

Qualitatively, we may summarize by saying that just a few $\tau$ (and possibly a fraction) suffice to achieve multidimensional behavior for $T\ge 0.25$ . For $T>15k_c\ell^6/32k_B$ a typical particle is able to pass through the position of another (the ``rigid balls" reduce to points).

We evaluated $\tau_p$ by counting bypassings. With the parameters considered above, we obtained  $\tau_p=0.31\tau$ for $T=1$ and $\tau_p=0.14\tau$ for $T=3.5$.

\subsubsection{Nyquist setup}
For the same wire studied in Fig.\ \ref{ImZw}, we evaluated the variances of the voltages and the emf's (in the absence of any power supply) between the ends of the entire wire ($0\le x\le L$), and the ends of segments of lengths $0.75 L$ ($0.25L\le x\le L$), $0.5L$ ($0\le x\le 0.5L$) and $0.25L$ ($0\le x\le 0.25L$), in the temperature range $0.25\le T\le 3.5$. The variances were obtained from $4\times 10^4$ measurements, each of duration $\Delta t=100\tau$, taken after a stabilization time of $6.5\times 10^5\tau$. Between measurements, an idle time of $30\tau$ was left to avoid correlations. The Drude formula was used without taking into account the deviations from uniform density.

\begin{figure}[tbh]
\scalebox{0.85}{\includegraphics{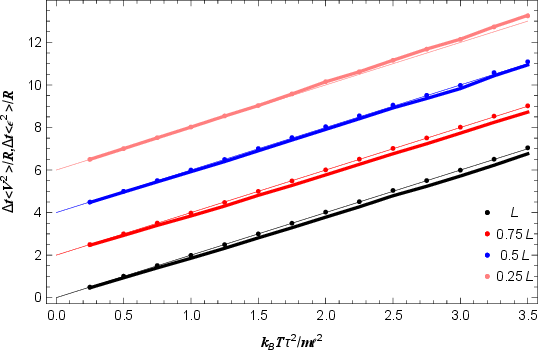}}
\caption{\label{unif2D}Variances of the emf and the voltage in an open circuit, as functions of the temperature. The straight thin dotted lines show the values predicted by Eq.\ (\ref{varep}), the prominent dots show our results for the emf, and the joined lines show the voltage variances. 
For visibility, the results for different segments have been shifted upwards in steps of 2 units.  Parameters: $N=80$, $L=20\ell$, $W=2.2\ell$, $k_c=15m\ell^{-4}\tau^{-2}$, $\Delta t=100\tau$.}
\end{figure}

Our results are presented in Fig.\ \ref{unif2D}. The dots show the variances of the emf, the thick lines show the variances of the voltages, and the dotted lines correspond to Eq.\ (\ref{varep}). Except for the shortest segment, there is good agreement between our results and Nyquist's prediction. Since electroneutrality is not perfect, the current is not negligible and the variance of the voltage differs from that of the emf. On thermodynamic grounds, the current and the voltage must be uncorrelated and therefore $\langle\epsilon^2\rangle = \langle V^2\rangle+ R^2\langle I^2\rangle >\langle V^2\rangle$. 
In view of these results, in the following we will not consider segments with less than 40 particles on average.

\begin{table*}
\caption{\label{tab:shield} Verification of Eq.\ (\ref{varep}) for a segment of  a wire that excludes the regions of length $L'=2.5\ell$ at the ends. Columns 2 and 3
(4 and 5) refer to temperature $T_1=1$ ($T_2=3.5$), $\epsilon_i$ is the emf and $R_i$ is the time average resistance of the segment when the temperature is $T_i$. The kinetic temperature $T_{k,i}$ is the average 2D kinetic energy of the particles in the segment divided by $k_B$.}
\begin{ruledtabular}
\begin{tabular}{c@{\hspace{1em}}cc@{\hspace{1em}}cc}
$N_{1,2}$ & $\Delta t \langle\epsilon_1^2\rangle /2R_1$  & $T_{k,1}$ &   $\Delta t \langle\epsilon_2^2\rangle /2R_2$  & $T_{k,2}$ \\
\hline
40 & 1.00 & 1.00 & 3.52 & 3.51 \\
60 & 1.01 & 1.00 & 3.49 & 3.51 \\
100 & 1.00 & 1.00 & 3.51 & 3.51 \\
150 & 1.01 & 1.00 & 3.52 & 3.51 \\ 
200 & 1.00 & 1.00 & 3.54 & 3.51 \\
250 & 1.00 & 1.00 & 3.52 & 3.51 \\
300 & 1.00 & 1.00 & 3.52 & 3.51 \\
350 & 1.01 & 1.00 & 3.54 & 3.51
\end{tabular}
\end{ruledtabular}
\end{table*}

Table \ref{tab:shield} shows the variances obtained for the emf in a wire, for a wide range of lengths, with fixed average linear density $4/\ell$, width $W=2.2\ell$, and force coefficient $k_c=15$. To minimize boundary effects, quantities were not evaluated in the range $0\le x\le L$, but rather in the range $L'\le x\le L-L'$, with $L'=2.5\ell$. This time we did not assume uniform or fixed density; we used instead Eq.\ (\ref{nonDrude}), where $n(x)$ was obtained by counting the number of particles in segments of length $\ell/16$ and averaging over the measurement duration $\Delta t$.
The number of particles in the wire is denoted by $N_{1,2}$ (rather than just $N$) with the intention of future comparison with a pair of coupled wires.

Columns 2 and 3 in Table \ref{tab:shield} refer to the temperature $T_1=1$, whereas columns 4 and 5 refer to $T_2=3.5$. $T_{k,i}$ is the average kinetic particle for temperature $T_i$; we see that $T_{k,i}>T_i$, i.e.\ the average kinetic energy of the particles is slightly larger than the energy imparted to them by the collisions. This extra energy is due to inaccuracy of the evolution algorithm; $T_{k,i}-T_i$ increases with the size of the evolution steps and with $T_i$. The Johnson temperature, $T_{J,i}:=\Delta t \langle\epsilon_i^2\rangle /2R_i$, is generally larger and has more statistical uncertainty than $T_{k,i}$; the extra variance may be attributed to the variance of the resistance \cite{Clarke}. For $N_{1,2}=40$ ($N_{1,2}=350$), denoting variance by $S$, $S(R_1)/R_1^2=1.7\times 10^{-4}$ and $S(R_2)/R_2^2=2.3\times 10^{-4}$ ($S(R_1)/R_1^2=1.2\times 10^{-5}$ and $S(R_2)/R_2^2=2.1\times 10^{-5}$).

\subsection{Capacitors}
In subsequent Sections we will consider contact between wires at different temperatures. We want these wires to be kept at uniform temperatures; accordingly, passage of particles between wires should be avoided. The obvious circuit element that enables passage of current while impeding passage of particles is a capacitor, and we therefore require a model for it.

Let us model a capacitor located at $x=x_o$, between wire 1 and wire 2. The region $|x-x_0|<L'$ will be regarded as a ``capacitor". We define the charges $Q_{1,2}$ on each side of the capacitor using a weighted count:
\begin{equation}
    Q_i = q \sum_{0<\sigma_i(x_j-x_0)< L'} \left(1-\frac{|x_j-x_0|}{L'}\right) - \frac{qNL'}{2L} 
\end{equation}
with $\sigma_1=-1$, $\sigma_2=1$, and $x_j$ the position of particle $j$, ensuring a smooth and local definition of $Q_{1,2}$. The electrostatic energy stored in the capacitor is 
\begin{equation}
    U=Q^2/2C+Q_T^2/2C' \,,
\end{equation}
where $Q=(Q_1-Q_2)/2$ is the usual ``capacitor charge", $Q_T=Q_1+Q_2$ is the total charge, $C$ is the mutual capacitance, and $C'$ is the stray (common-mode) capacitance. Therefore, the force exerted by the capacitor on a particle $j$ on side $i$, within a distance $L'$ from the interface, is
\begin{equation}
    -\frac{\partial U}{\partial x_j}=-\frac{qQ}{2L' C}+\sigma_i\frac{qQ_T}{L' C'} \,.
\end{equation}
We note that this force acts in addition to the other forces (that do not depend on $Q_i$) exerted on particle $j$.
We will take $C\to\infty$ and small $C'$. By doing this, the capacitor should have no impedance and should aim to establish continuity of the current. Since in reality the force exerted by the capacitor and the inter-particle force are both electrostatic, a natural requirement is $C'\propto k_c^{-1}$. 

As a test for the behavior of a capacitor in a circuit, we evaluated again the impedance between the ends of the wire considered in Sec.\ \ref{Ohm}, this time with a capacitor inserted in the middle of the wire (i.e.\ centered at $x=L/2$). If the model capacitor does behave as a standard capacitor with infinite capacitance, it should add no impedance to the wire, for any nonzero frequency. Our results are shown by the green line in Fig.\ \ref{ImZw}. Re[$Z$] is close to our previous results (that were independent of the temperature), but Im[$Z$] shows deviations. The main reason for these deviations is that splitting a wire of length $L=20\ell$ results in halves that are too short to neglect end effects; taking a wire of double length yields results consistent with the case in which there is no capacitor.

\section{Circuit with resistors at different temperatures}
\begin{figure}[tbh]
\scalebox{0.85}{\includegraphics{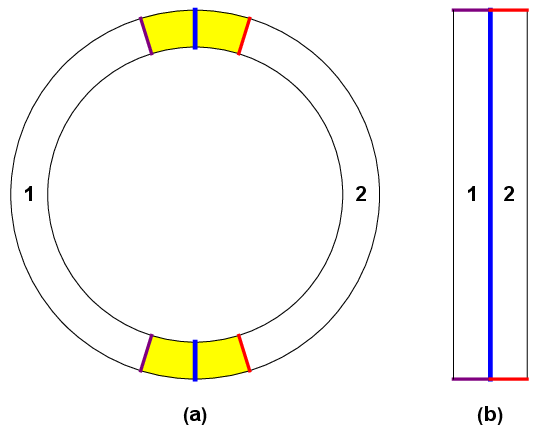}}
\caption{\label{configurations}Configurations investigated in this article. 
Wire 1 (2) has $N_1$ ($N_2$) particles, length $L_1$ ($L_2=L-L_1$) and temperature $T_1$ ($T_2$). The wires make contact with each other at the thick blue lines.
The yellow regions are regarded as capacitors. (a) End-to-end configuration; (b) side-by-side configuration.}
\end{figure}

We study here the end-to-end configuration shown in Fig.\ \ref{configurations}(a). The side-by-side configuration, shown in Fig.\ \ref{configurations}(b), will be considered in Appendix \ref{SbS}.

We revisit the examples reported in Table \ref{tab:shield}, but now we connect pairs of wires by means of two capacitors, centered at $x=L_1$ and $x=L$ (equivalent to $x=0$), as shown in Fig.\ \ref{configurations}(a). $V_1$ and $\epsilon_1$ ($V_2$ and $\epsilon_2$) are measured between the black (red) thick lines in wire 1 (wire 2). As in Table \ref{tab:shield}, $T_1=1$, $T_2=3.5$, and $N_1/L_1=N_2/L_2=4$. The capacitors extend to a distance $L'=2.5\ell$ from the contact lines. Our results are presented in Table \ref{tab:cap}.

\begin{table*}
\caption{\label{tab:cap} Variances of the voltages, changes of the Johnson and kinetic temperatures, and inter-correlations in a pair of wires connected by capacitors as in Fig.\ \ref{configurations}(a). In all cases $T_1=1$, $T_2=3.5$, the average density of particles is $4/\ell$ in each wire, $k_c=15$, $C'=1/k_c$ and segments of length $L'=2.5$ are excluded from the measurements. Not all the reported digits are free of statistical error.}
\begin{ruledtabular}
\begin{tabular}{c@{\hspace{1em}}ccc@{\hspace{1em}}ccc@{\hspace{1em}}cc}
$N_{1,2}$  & $\frac{\Delta t \langle V_1^2\rangle }{2R_1}$ & $\Delta T_{J,1}$ & $\Delta T_{k,1}$ & $\frac{\Delta t \langle V_2^2\rangle }{2R_2}$ & $\Delta T_{J,2}$ & $\Delta T_{k,2}$& $r(\epsilon_1,\epsilon_2)$ & $r(V_1,V_2)$ \\
\hline
40 & 0.97 & 0.027 & 0.018 & 3.41 & -0.034 & -0.024 & -0.003 & -0.014 \\
60 & 0.92 & 0.029 & 0.009 & 3.26 & -0.045 & -0.012 & -0.002 & -0.020  \\
100 & 0.75 & 0.030 & 0.005 & 2.91 & -0.039 & -0.006 & -0.006 & -0.048 \\
150 & 0.56 & 0.033 & 0.003 & 2.30 & -0.045 & -0.004 & -0.014 & -0.082 \\
200 & 0.41 & 0.021 & 0.002 & 1.80 & -0.042 & -0.002 & -0.007 & -0.084  \\
250 & 0.32 & 0.018 & 0.002 & 1.42 & -0.036 & -0.002 & -0.002 & -0.083 \\
300 & 0.26 & 0.013 & 0.001 & 1.17 & -0.015 & -0.002 & -0.006 & -0.083 \\
350 & 0.22 & 0.002 & 0.001 & 1.00 & -0.016 & -0.001 & -0.001 & -0.076
\end{tabular}
\end{ruledtabular}
\end{table*}

\subsection{Effective Johnson temperatures}

The variance of the voltage is not directly related to Nyquist's formula but is given in the Table since it, and not the emf, is directly measurable. $\langle V_i^2\rangle$ is smaller than $\langle\epsilon_i^2\rangle$, as it should; the share of the variance taken by the current fluctuations, $R_i^2\langle I_i^2\rangle=\langle\epsilon_i^2\rangle -\langle V_i^2\rangle$, increases with the length of the wires.

We denote by $T_{J,i}^{\rm disc}$ the Johnson temperature in wire $i$ when it is not connected to the other wire, by $T_{J,i}^{\rm cap}$ the Johnson temperature in wire $i$ when connected by means of capacitors, and by $\Delta T_{J,i}:=T_{J,i}^{\rm cap}-T_{J,i}^{\rm disc}$ the increase due to the connection. Similarly, we denote by $\Delta T_{k,i}:=T_{k,i}^{\rm cap}-T_{k,i}^{\rm disc}$ the increase in kinetic temperature due to the connection. Since (with the exception of the short regions occupied by the capacitors) the same forces, algorithm, and initial state were used when the wires were connected and when they were disconnected, systematic inaccuracies are expected to cancel out in  $\Delta T_{J,i}$ and $\Delta T_{k,i}$.

We see in Table \ref{tab:cap} that $\Delta T_{J,1}$ is systematically positive, namely, taking the disconnected situation as the baseline, when a wire at uniform temperature is connected to another wire at higher temperature, the Johnson noise in the former is higher than predicted by the Nyquist relation. The analogous situation occurs in wire 2. This is the central result of this article: the variance of the emf in a resistor with local temperature $T_i$ is not determined solely by this local temperature, but is influenced by the temperatures in other resistors provided that they have electrical contact. Moreover, $\Delta T_{J,1}>\Delta T_{k,1}$ and $|\Delta T_{J,2}|>|\Delta T_{k,2}|$, so the change in Johnson noise cannot be explained by the difference between the temperatures of the particles and those of the lattices that emerge from the electrical contact. 

\begin{figure}[tbh]
\scalebox{0.85}{\includegraphics{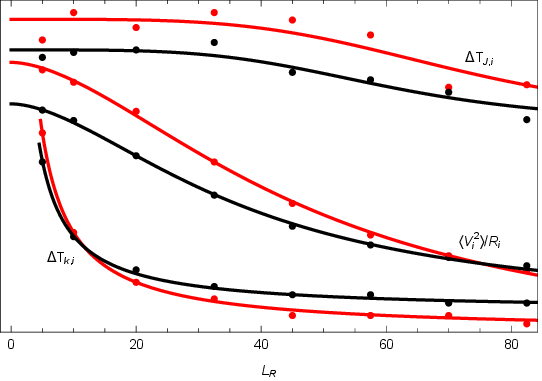}}
\caption{\label{visualII} Qualitative visualization of the results given in Table \ref{tab:cap}. Different lines in the graph generally have different scales and vertical position. $L_R=L_{1,2}-2L'$ is the length of the segment over which measurements are taken. Red lines and dots refer to $T=T_2=3.5$; black refers to $T=T_1=1$, with fitting values given in brackets. Fitting lines: $|\Delta T_{J,i}|=A/[1+(L_R/\lambda)^4]$, $A=0.042$, $\lambda =73$ ($A=0.030$, $\lambda =61$); $\Delta t \langle V_i^2\rangle /2R_i=T_i/[1+(L_R/\lambda)^{1.74}]$, $i=2$, $\lambda =47$ ($i=1$, $\lambda =38$); $|\Delta T_{k,i}|=A/L_R$, $A=0.12$ ($A=0.09$).}
\end{figure}

Figure \ref{visualII} visualizes the dependence of the results in columns 2-7 of Table \ref{tab:cap} on the reduced length $L_R=N_{1,2}\ell /4-2L'$. The fits are empiric. The fits for $\Delta T_{k,i}$ can be interpreted as follows: the capacitor passes power from the hot to the cold particles and this power is independent of $L_R$. Then this power is transferred to the lattice, and is proportional to $T_{k,i}-T_i$ and to $L_R$ (even if the kinetic energy density is not uniform). Therefore, $|T_{k,i}-T_i|$ is inversely proportional to $L_R$. If systematic errors are not neglected and the values for separate wires are taken as the baseline, $T_{k,i}-T_i$ has to be replaced by $\Delta T_{k,i}$.

The results for $\Delta T_{J,i}$ are scattered, but the fit captures their qualitative behavior. $\Delta T_{J,i}$ does not change significantly for $L_R\alt\lambda_i\sim 60\ell$, but decreases abruptly for $L_R\agt\lambda_i$. This dependence on the length is very different from that of $\Delta T_{k,i}$, supporting the view that noise transfer is not just a consequence of energy transfer. For $k_c\to 0$ the wires become decoupled and therefore $\Delta T_{k,i}\to 0$, but $\Delta T_{J,i}$ only slightly yields to variation of $k_c$; reducing $k_c$ from 15 to 2 and keeping the fit $|\Delta T_{J,i}|=A_i/[1+(L_R/\lambda_i)^4]$ results in the decrease of $\lambda_{1,2}$ by roughly 40\% and a moderate increase of $A_{1,2}$. 

\subsection{Correlations}
We denote the correlation between random variables $X_1$ and $X_2$ by $r(X_1,X_2):=\langle (X_1-\langle X_1\rangle)(X_2-\langle X_2\rangle)/\sqrt{S(X_1)S(X_2)}\rangle$. Accepting the view that the emf is due to local thermal agitation [the Nyquist scenario (NS)] implies that the emf's in two detached segments should be uncorrelated.
The results in Table \ref{tab:cap} disagree with this prediction. In this Section we explore this question further, this time, motivated by Eq.\ (B6) in \cite{BK}, for wires of different length.

Since the emf is not directly measurable, let us translate the NS condition into a condition for voltages.We first consider the case that Kirchhoff's incompressibility assumption is obeyed.Then, 
from (\ref{defep}), the NS condition becomes
\begin{equation}
    \langle V_1V_2\rangle+R_1\langle V_2I\rangle +R_2\langle V_1I\rangle +R_1R_2\langle I^2\rangle =0 \,,
    \label{N1}
\end{equation}
where the resistances $R_{1,2}$ have been taken as constants.

From (\ref{varep}) and (\ref{defep}),
\begin{equation}
  \langle V_i^2\rangle+2R_i\langle V_iI\rangle+R_i^2\langle I^2\rangle=2k_BT_iR_i/\Delta t \,.
  \label{N2}
\end{equation}
Eliminating in (\ref{N1}) and (\ref{N2}) $\langle V_1I\rangle$ and $\langle V_2I\rangle$ (which do not vanish for $T_1\neq T_2$), we obtain
\begin{equation}
   \langle (R_1V_2-R_2V_1)^2\rangle =2R_1R_2k_B(R_1T_2+R_2T_1)/\Delta t 
   \label{riVi}
\end{equation}
(regardless of $\langle I^2\rangle$).

Equation (\ref{riVi}) may be useful in a real experiment, but in our simulations incompressibility is difficult to achieve. In Table \ref{tab:compress} we consider two wires in the configuration of Fig.\ \ref{configurations}(a), with $W=1.5$, $k_c=50$, $N_1=40$, $N_2=80$, and nominal densities $N/L=4$. These values are beyond our preferred choices, but the purpose here is to approach incompressibility. 
Despite this choice, we see in Table \ref{tab:compress} that $r(I_1 I_2)$ is far from 1, as would follow from Kirchoff's law; nevertheless, the last column in the Table is not far from 1, as required from Eq.\ (\ref{riVi}). $r(\epsilon_1,\epsilon_2)$ changes sign between the cases $T_1\gg T_2$ and $T_1\ll T_2$; for $T_1=T_2$, $|r(\epsilon_1,\epsilon_2)|$ is quite smaller and could be due to insufficient accuracy of our evaluations.

\begin{table*}
\caption{\label{tab:compress} Correlation of the emf's and related quantities for a pair of wires coupled by capacitors as indicated in Fig.\ \ref{configurations}(a). $k_c=50$, $W=1.5$, $L_1=10$ and $L_2=20$ (see text for additional parameters). The segments at the ends of the wires that act as capacitors are not included in the evaluated quantities.}
\begin{ruledtabular}
\begin{tabular}{ccccc}
$T_1$ & $T2$ & $r(\epsilon_1,\epsilon_2)$ &$r(I_1,I_2)$ & $ \frac{\Delta t\langle (R_1V_2-R_2V_1)^2\rangle }{2R_1R_2k_B(R_1T_2+R_2T_1)}$ \\
\hline
 3.2 & 0.8 & -0.009 & 0.22 & 0.994 \\
 2.0 & 2.0 & -0.002 & 0.29 & 0.996 \\
 0.8 & 3.2 & ~0.008 & 0.40 & 0.990
\end{tabular}
\end{ruledtabular}
\end{table*}

\begin{figure}[tbh]
\scalebox{0.85}{\includegraphics{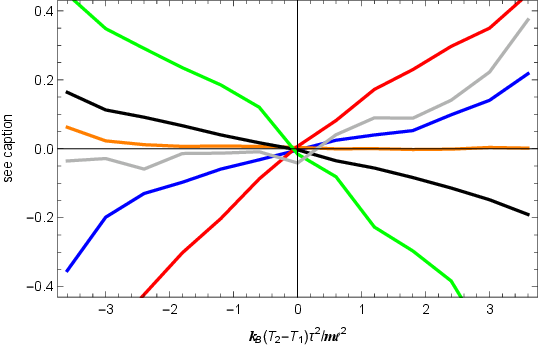}}
\caption{\label{figcorrel} Quantities that arise from the capacitive coupling between a pair of wires connected as in Fig.\ \ref{configurations}(a), as functions of their temperature difference. The sum of temperatures is kept fixed, $T_1+T_2=4$. $k_c=15$, $W=2.2$, $L_1=25$ and $L_2=50$. 
Blue: $3r(\epsilon_1,\epsilon_2)$; orange: $r(V_2,I_2)$; black: $\Delta t\langle (R_1V_2-R_2V_1)^2\rangle /2R_1R_2k_B(R_1T_2+R_2T_1)-0.7-0.3r(I_1,I_2)$; red: $E_{kx1}$ ($E_{kx1}[-3.6]=-0.68$); green: $E_{kx2}$ ($E_{kx2}[3.6]=-0.67$); gray: $3E_{ky1}$.
The segments at the ends of the wires that act as capacitors are not included in the evaluated quantities. }
\end{figure}

For a more detailed study, and for comparison with Table \ref{tab:cap}, we took the parameters $W=2.2$, $k_c=15$, $N_1=100$, and $N_2=200$.
The wire temperatures were varied, but their average was kept constant at $(T_1+T_2)/2=2$. The results are presented in Fig.\ \ref{figcorrel}.
 Our central target value, $r(\epsilon_1,\epsilon_2)$, is given by the blue line. As in the 1D case \cite{BK}, we find that $r(\epsilon_1,\epsilon_2)$ is positive (negative) if the longer wire is hotter (colder) than the shorter wire. As in the 1D case, the slope of this line increases as $|T_2-T_1|$ increases. $r(\epsilon_1,\epsilon_2)$ is not an odd function; for given $\Delta T=T_2-T_1$, $r(\epsilon_1,\epsilon_2)[\Delta T]+r(\epsilon_1,\epsilon_2)[-\Delta T]<0$. Also in the degenerate case of wires of the same length presented in Table \ref{tab:cap}, negative values of $r(\epsilon_1,\epsilon_2)$ prevail. 

Since only longitudinal motion is relevant for the work performed by the capacitor, we separate the contributions to the kinetic energy from the $x$- and the $y$-components of the velocities. We denote by $E_{kx1}$ (and analogously $E_{kx2}$, $E_{ky1}$ and $E_{ky2}$) the kinetic energy in wire 1 (excluding the capacitor regions), due to the $x$-components of the velocities, additional to the average kinetic energy $\langle N'_1\rangle k_BT_1/2$, where $\langle N'_1\rangle$ is the average number of particles in the considered segment. If this additional energy builds up at uniform rate, the contribution of $E_{kx1}$ to the power transferred from the particles to the lattice in wire 1 is $2E_{kx1}/\tau$. $E_{kx1}$ ($E_{kx2}$) is described by the red (green) line in Fig.\ \ref{figcorrel}. Despite the large ratio $(L_2-2L')/(L_1-2L')>2$, there is almost symmetry between the line for $E_{kx1}$ and the line for $E_{kx2}$, indicating that most of the particles with energies that deviate from the average are located close to the capacitor. We also evaluated the excess kinetic energy per degree of freedom, $(m<v_x^2>-k_BT)/2$, as a function of position. As expected, $|m<v_x^2>-k_BT|/2$ is largest next to the contact surfaces between the two wires; it decreases by an order of magnitude at a distance of 3-4$\ell$ from these surfaces. Within the capacitors and for low temperatures, $|m<v_y^2>-k_BT|$ is smaller than $|m<v_x^2>-k_BT|$ by an order of magnitude. $E_{ky1}$ is given by the gray line.

The orange line in Fig.\ \ref{figcorrel} stands for $r(V_2,I_2)$. For uniform current or ``electric" field, $\langle V_2 I_2\rangle$ is the average power delivered by the voltage to the particles in wire 2. For $T_2-T_1\alt -3$, $r(V_2,I_2)$ is significant, but even in this range $\langle V_2 I_2\rangle$ is smaller than $E_{K,xi}/\tau$ by an order of magnitude, marginalizing this macroscopic mechanism as the process by which energy is transferred from the hot to the cold wire.

We denote $\chi :=\Delta t\langle (R_1V_2-R_2V_1)^2\rangle /2R_1R_2k_B(R_1T_2+R_2T_1)-1+\alpha [1-r(I_1,I_2)]$. If incompressibility holds, $r(I_1,I_2)$ equals 1 and, therefore, Eq.\ (\ref{riVi}) is equivalent to $\chi =0$ for arbitrary $\alpha$; if $r(I_1,I_2)\neq 1$, we adjust $\alpha$ to obtain $\chi =0$ at $T_1=T_2$. By continuity, we expect that this choice gives a rough correction to the expression with $\alpha =0$. $\chi$ is described by the black line in Fig.\ \ref{figcorrel}. 

By detailed balance (or by construction), all the curves in Fig.\ \ref{figcorrel} should pass through the origin; the deviations are due mainly to statistical uncertainty.


\section{Summary and notes}
A previous article \cite{BK} raised the claim that thermal noise (the emf) in a resistor depends not only on its own temperature, but also on the temperatures of the other resistors in the circuit. Furthermore, it was claimed that the strong repulsion between charges, via Kirchhoff's incompressibility law, is the agent that transmits the influence between the resistors.

As a test of this claim, here we performed simulations with systems of classical particles with limited-range repulsive interaction. Wires were mimicked by rectangles of lengths larger than the widths by one or two orders of magnitude. The velocities of all the particles were regularly and simultaneously thermalized by a heat bath. Although experiments are expected to isolate a frequency band to improve signal to noise ratio, it was easier to express our results in the time domain. The simplicity of our model allowed us to follow the evolution of systems with several hundred particles over times during which every particle underwent more than $10^7$ collisions. In this modified Drude model the resistance has twice the value and the kinetic inductance has two thirds the value that they would have in the ``standard" Drude model (where the collision probability during any infinitesimal time $dt$ is $dt/\tau$).

Pairs of wires were not connected directly; instead they were coupled via ideal capacitors. These capacitors had infinite ``usual" (mutual) capacity and small stray capacity (required to obtain nearly uniform current density). The force exerted by the capacitors on the particles was inspired by electrostatics and was a smooth function of position.

The present study is ``cleaner" than that in \cite{BK} in several aspects: (1) multidimensionality avoids the correlations that arise when particles are forced to keep the same sequential order during long periods of time; (2) it focuses on emf's (as in the Nyquist scenario) and voltages (easily measurable) rather than currents; (3) there is no leakage of particles between resistors; (4) we do not need to consider regions where the temperature is not uniform. With the parameters we used, our model fails to give a good approximation to Kirchhoff's law; we therefore hold that experiments will find influence between resistors at distances that are longer than predicted by our simulations.

As predicted in \cite{BK}, when two resistors at different temperatures are connected in series [more precisely, as in Fig.\ \ref{configurations}(a)], their Johnson temperatures approach that of the other resistor. The effect is not dramatic as in the 1D case; the shift of the Johnson temperatures is smaller than the temperature difference between the resistors by two orders of magnitude, and extends to distances that are longer than the interaction range between the particles by two orders of magnitude. The emf's of the two resistors may be correlated by several percent, especially if the two wires have different lengths.

We  studied in an Appendix the case of adjacent wires as in Fig.\ \ref{configurations}(b). The purpose of this study was to gain some understanding of what to expect in the case of a multidimensional resistor (not necessarily narrow in any direction) with nonunuform temperature. Again, we found that the Johnson temperature in each of the wires is influenced by the temperature in the other wire, but this effect is obscured by heat transfer.

The propagation speed of Johnson noise remains an open question. We conjecture that the motion responsible for the propagation of Johnson
 noise is acoustic-like and may therefore exhibit behavior fundamentally different from that of a diffusive process. Preliminary nonstationary simulations, in which the two wires were intermittently coupled, revealed that the switching process itself induces significant correlations and pseudo-Johnson noise, even when both wires are at the same temperature. A quantitative investigation of this effect is therefore left for future work.

Our model and Rytov's theory \cite{Ry} differ in scope: we consider only the quasistatic regime in which there is no radiation, we do not invoke the macroscopic local Ohm's law, statistical independence is imposed on the particles' velocities after collisions rather than on the current density sources, the electric field that couples the wires is not obtained from Maxwell's equations, but as an electrostatic field that depends only on the amount of charge in the capacitor, and noise propagation is mediated by a presumably acoustic-like process rather than by an electromagnetic wave.

We conclude this summary by comparing the correlation $r(\epsilon_1,\epsilon_2)$ shown in Fig.\ \ref{figcorrel} (or the analytic 1D analog, described by Eq.\ (B6) in \cite{BK}) with the results derived from the fluctuation-dissipation theorem, considered in Appendix \ref{FDT}. In equilibrium, it follows from FDT that the fluctuating longitudinal electric field $E(x,t)$ obeys
\begin{equation}
    \langle E(x,t) E(x',t') \rangle = 2 k_B T\rho (x) \delta(x - x') \delta(t - t') \,,
    \label{localcorr}
\end{equation}
where $\rho$ is the resistance per unit length, implying that the emf's in disjoint segments are uncorrelated. The assumptions that underlie Eq. (\ref{localcorr}) are exposed in Appendix \ref{FDT}. 
In local-Langevin and Boltzmann-Langevin descriptions, the stochastic sources are taken to be local, i.e. proportional to $\delta (x-x')$, also in nonequilibrium formulations. Physically, if the typical displacement $\delta x$ of a particle until it re-thermalizes obeys $T(x+\delta x)\approx T(x)$, then it is justified to assume that there is local thermodynamic equilibrium \cite{hydro,Maz}, and relations analogous to Eq.\ (\ref{localcorr}) are commonly used with the temperature $T$ replaced with a local temperature $T(x)$, e.g. Chapter 3 in \cite{hydro} or \cite{Henkel}.

In the present study, for the equilibrium situation $T_1=T_2$, $\epsilon_1$ and $\epsilon_2$ are uncorrelated, in agreement with FDT. For large values of $T_1/T_2$ or $T_2/T_1$ we obtained correlations of the order of 10\%, in disagreement with the assumption of independent local noise sources. We conjecture that this disagreement originates from the combination of nonzero-range particle interactions and approximate compliance with Kirchhoff's law. We recall that the segments in which $\epsilon_1$ and $\epsilon_2$ were evaluated are detached: the minimum distance between any pair of parts in them is $5\ell$ and the distance between their centers is $37.5\ell$; the only coupling between the two wires is the electrostatic interaction transmitted through the capacitors.

\begin{acknowledgments}
The author has benefited from correspondence with Carsten Henkel, Guy Katriel, Jan Sengers and Boris Shapiro, and from consults with ChatGPT, Claude and Gemini.
\end{acknowledgments}
\appendix


\section{Impedance in a modified Drude model with synchronous collisions }
The standard Drude model assumes that collisions occur as a Poisson process. In the present work, we instead assume that all particles undergo thermalizing collisions simultaneously at equally spaced time intervals. This modification simplifies the numerical implementation while preserving the transport properties relevant to the present study. 
We study a system of $N$ particles with charge $q$ and mass $m$ in a wire of length $L$. This Drude model requires that, at every time $t=n\tau$ with $n$ integer, the average velocity be set to zero. In this Appendix we ignore thermal fluctuations, interactions between particles, and averaging during measurement durations.

Let there be a uniform field $E\exp (i\omega t)$ in the wire. The voltage will be $V=EL\exp (i\omega t)$ and the acceleration of every particle will be $(qE/m)\exp (i\omega t)$. Integrating, the velocity during $n\tau\le t <(n+1)\tau$, averaged over particles, will be $(qE/i\omega m)[\exp (i\omega t)-\exp (i\omega n\tau)]$ and, using (\ref{defI}), the current is $I(t)=(Nq^2E/i\omega mL)[\exp (i\omega t)-\exp (i\omega n\tau)]$.

Consider a period of time $M\tau$ with integer $M$ such that, to a good approximation, $\omega$ is an integer multiple of $2\pi/M\tau$. Then $I(t)$ can be taken as periodic with period $ M\tau$ and the Fourier series of $I(t)$ contains a term of the form $I_\omega\exp (i\omega t)$. The coefficient $I_\omega$ can be obtained as $I_\omega=(1/M\tau)\int_0^{M\tau}I(t)\exp (-i\omega t)dt$. Splitting into segments, we obtain
\begin{equation}
    I_\omega=\frac{Nq^2E}{i\omega mLM\tau}\left[ \int_0^{M\tau} 1 \, dt -\sum_{n=0}^{M-1} \int_{n\tau}^{(n+1)\tau} e^{i\omega n \tau} e^{-i\omega t} dt\right]\;.
\end{equation}
The first integral is $M\tau$ and each integral within the sum is $(\exp[-i\omega\tau]-1)/(-i\omega)$. Therefore,
\begin{equation}
    I_\omega=\frac{Nq^2E}{i\omega mL}\left[1-\frac{\exp[-i\omega\tau]-1}{-i\omega\tau} \right]= 
    \frac{Nq^2E}{\omega^2 mL\tau}(1-\exp[-i\omega\tau]-i\omega\tau)\;.
\end{equation}

Dividing $I_\omega$ by the amplitude of the voltage and comparing with (\ref{Drude}) we obtain the admittance
\begin{equation}
    \frac{1}{Z_\omega}=\frac{2(1-\exp[-i\omega\tau]-i\omega\tau)}{R\omega^2\tau^2}\;.
    \label{Y}
\end{equation}
In the limit $\omega\tau\to 0$, we obtain $Z_0=R$, as required. Unlike the impedance in the ``standard'' Drude model, $R(1+i\omega\tau)$, (\ref{Y}) contains an oscillatory term that arises from the sharp distribution of the collision times. In the region in which we are interested, $\omega\tau\alt 1$, oscillations don't show up. Expanding in powers of $\omega\tau$ we obtain
\begin{equation}
    \frac{\mathrm{Re}Z_\omega}{R}=1-\frac{\omega^2\tau^2}{36}-\frac{\omega^4\tau^4}{6480}+O[\omega^6\tau^6]
\end{equation}
and
\begin{equation}
    \frac{\mathrm{Im}Z_\omega}{R}=\frac{\omega\tau}{3}+\frac{\omega^3\tau^3}{540}+\frac{\omega^5\tau^5}{27216}+O[\omega^7\tau^7] \;.
    \label{ImZ}
\end{equation}
For $\omega\tau =0.5$, $\mathrm{Re}Z_\omega$ is 0.7\% smaller than $R$. It should be noted that the kinetic inductance \cite{Lk} is already included in (\ref{ImZ}) and should not be added.

\section{Side-by-side configuration \label{SbS}}
We consider here the pair of side-by-side wires, each with uniform temperature $T_1$ or $T_2$, illustrated in Fig.\ \ref{configurations}(b). Between the wires we envisage an infinitesimal layer perfectly impermeable to the passage of particles or phonons. There is no clear electric connection between them, but rather a sort of distributed parasitic capacitance. 
A similar situation could be that of two  conductors that exchange black-body radiation \cite{Henkel,Abdalla}.
\subsection{Model}
The model for each wire will be the same as for an isolated wire, except that along the common lateral wall, instead of repulsion from image particles, there will be interaction with the particles on the other side of the wall, as if there were no wall. The only role of the wall is to reflect the particles that reach it. The current along a wire will still be given by Eq.\ (\ref{defI}) and the voltage by Eq.\ (\ref{defV}), where now the interactions with the particles on the other side of the wall are included as external forces.

\subsection{The Nyquist scenario}
Application of Nyquist's formula is complicated by the inter-wire interaction, which behaves as a non-obvious circuit element. The prediction for the emf's is immediate: these are local quantities and should be given by Eq.\ (\ref{varep}), as if there were no interaction.

In the absence of a clear equivalent circuit for the inter-wire interaction, we investigated the correlations $r(\epsilon_1,\epsilon_2)$, $r(V_1,V_2)$ and $r(I_1,I_2)$ for $T_1=T_2$, in which case this interaction does not compete with nonlocality. As in previous sections, we took the widths $W=2.2\ell$, average linear densities $N/L=4/\ell$, and force constant $k_c=15$, together with $T_{1,2}$ 1 or 3.5, $N_{1,2}$ 40 or 150 and $\Delta t$ 33 or 100. As expected, $r(\epsilon_1,\epsilon_2)$ vanished within statistical uncertainty, but this was also the case for $r(V_1,V_2)$. However, we found $r(I_1,I_2)<0$; the most negative value, $r(I_1,I_2)\approx -0.1$, was obtained for $T_{1,2}=1$, $N_{1,2}=40$ and $\Delta t=33$.

The result $r(I_1,I_2)<0$ suggests that the current that cannot continue past an end of a wire has some tendency to return along the other wire. The result $r(V_1,V_2)\sim 0$ gives some confidence that, if for $T_1\neq T_2$ some experiment finds that the voltage variance along a wire depends on the noise along its neighbor, or if the two voltages are correlated, then the same holds for the emf's.
 
\subsection{Results}
Table \ref{tab:side} is the counterpart of Table \ref{tab:cap} for the side-by-side situation. 
For pairs of long wires, the voltage variances are slightly smaller than in the end-to-end case. We attribute this difference to the ease of current passage between the wires; the easier the passage, the larger the currents along both wires, the larger their variances, and thus the smaller the voltage variances. 
Unlike the end-to-end case, the size of the contact region between the wires is not fixed, but proportional to the wires length. Therefore, we expect that $\Delta T_{k,1}$ and $\Delta T_{k,2}$ (previously inversely proportional to the length) will be independent of the length, whereas $\Delta T_{J,1}$ and $\Delta T_{J,2}$ (roughly length independent for short wires in the previous case) will be proportional to the length within the same range. Our results qualitatively confirm this expectation. 
 The correlations $r(\epsilon_1,\epsilon_2)$ have the opposite sign and are more erratic than in the end-to-end configuration. Also in $r(V_1,V_2)$ we see differences, that may be more influenced by circuit properties than by nonlocality.
 We suspected that the energy exchange between the wires could be a source of anisotropy and therefore evaluated separately the contributions of $v_x^2$ and $v_y^2$ to $\Delta T_k$, but no significant anisotropy was found.

Since the dependence on the length of $\Delta T_{J,1}$ and $\Delta T_{J,2}$ is different from that of $\Delta T_{k,1}$ and $\Delta T_{k,2}$, we conclude that the dependence of the Johnson temperatures on the temperature of the neighboring wire is not due just to the energy exchange between them. However, in contrast to the end-to-end configuration, $|\Delta T_J|<|\Delta T_k|$ in the considered range.

\begin{table*}
\caption{\label{tab:side} Variances of the voltages, changes of the Johnson and kinetic temperatures, and inter-correlations in a pair of wires separated by an infinitesimal isolating layer as in Fig.\ \ref{configurations}(b). In all cases $T_1=1$, $T_2=3.5$, the average density of particles is $4/\ell$ in each wire, and $k_c=15$. Not all the reported digits are free of statistical error.}
\begin{ruledtabular}
\begin{tabular}{c@{\hspace{1em}}ccc@{\hspace{1em}}ccc@{\hspace{1em}}cc}
$N_{1,2}$  & $\frac{\Delta t \langle V_1^2\rangle }{2R_1}$ & $\Delta T_{J,1}$ & $\Delta T_{k,1}$ & $\frac{\Delta t \langle V_2^2\rangle }{2R_2}$ & $\Delta T_{J,2}$ & $\Delta T_{k,2}$& $r(\epsilon_1,\epsilon_2)$ & $r(V_1,V_2)$ \\
\hline
40 & 0.96 & -0.001 & 0.057 & 3.44 & -0.002 & -0.049 & 0.003 & 0.006 \\
100 & 0.75 & 0.021 & 0.058 & 2.94 & -0.001 & -0.051 & 0.009 & 0.022 \\
200 & 0.44 & 0.050 & 0.057 & 1.86 & -0.037 & -0.050 & 0.001 & 0.021  \\
300 & 0.30 & 0.053 & 0.057 & 1.25 & -0.048 & -0.051 & 0.007 & 0.009 
\end{tabular}
\end{ruledtabular}
\end{table*}

We also performed a superficial investigation of the case of a pair of wires with different widths, namely, for $L_1=L_2=25$, $W_2=2W_1=4.4$, $N_2/L_2=2N_1/L_1=8$, and $T_1+T2=4$. Again, we obtained $|\Delta T_J|\alt |\Delta T_k|$ and $r(\epsilon_1,\epsilon_2)\ge 0$. For $T_1=T_2$ we obtained the correlations $r(\epsilon_1,\epsilon_2)< 0.01$, $r(V_1,V_2)=0.02$, $r(I_1,I_2)=-0.04$ and, for $|T_2-T_1|=3.6$, $r(\epsilon_1,\epsilon_2)\approx 0.03$, $r(V_1,V_2)\approx 0.04$, $r(I_1,I_2)\approx -0.07$. 

\section{Fluctuation-dissipation theorem in a modified Drude model with synchronous collisions \label{FDT}}
\subsection{Single particle}
For a macroscopic analysis we introduce coarse-graining of time and consider a lapse of time $\delta t\gg\tau$ that will be taken as macroscopically infinitesimal. Let $F_e$ be an external force in the $x$-direction that can be taken as constant during $\delta t$, and let $\langle v_x\rangle(s)$ denote the ensemble-averaged velocity at time $s$ after a collision. Since the velocity is re-drawn from the equilibrium distribution at each collision, $\langle v_x\rangle(0)=0$; and since between collisions the only systematic force is $F_e$ (between collisions there is no systematic retarding force, because in this model friction is generated entirely by the resets), $\langle v_x\rangle(s) = F_e s/m$. The drift velocity, obtained by
averaging over $\delta t/\tau$ cycles, is therefore $\bar v_x = \tau^{-1}\int_0^\tau \langle v_x\rangle(s)\,ds = F_e\tau/2m$ and the friction coefficient is $\gamma=F_e/\bar v_x =2m/\tau$.

In equilibrium, for $\tau$ sufficiently small to allow neglecting the variation of $v_x$ during a cycle, the velocity autocorrelation is
\begin{equation}
    C(t)=\langle v_x(s)v_x(s+t)\rangle =\frac{k_BT}{m}\left(1-\frac{|t|}{\tau}\right) \;\;,
    |t|<\tau \;\; \text{(zero otherwise)},
    \label{Ct}
\end{equation}
where we have averaged over the phase $s$ within the collision cycle, and also over $v_x^2(s)$, setting $v_x^2=k_BT/m$. Green–Kubo integration gives the diffusion constant
\begin{equation}
    D=\int_0^\infty C(t)dt=\frac{k_BT\tau}{2m}=\frac{k_BT}{\gamma}\;,
\end{equation}
satisfying the Einstein-Smoluchowski relation.

To make contact with a Langevin description, and in the absence of external forces, we write the net force acting on the particle as $m\,\dot{ v}_x = -\gamma\,v_x + f_{\mathrm{int}}(t)$, where $f_{\mathrm{int}}$ is a Langevin stochastic force with vanishing equilibrium mean that stands for the rapidly varying part of the force exerted by the lattice and the other particles and is not accounted for in the term $ -\gamma\, v_x $. 

From $f_{\mathrm{int}} = m\dot v_x + \gamma v_x$, both terms on the right decorrelate on the collision time, hence so does $f_{\mathrm{int}}$,  and we therefore write $\langle f_{\mathrm{int}}(t)f_{\mathrm{int}}(t')\rangle
= A\,\delta_\tau(t-t')$, where $A$ is a constant and $\delta_\tau$ is a ``macroscopic $\delta$-function". By this we mean that $\delta_\tau$ vanishes if its argument is significantly larger than $\tau$, its total area is 1, and its precise shape is irrelevant for phenomena resolved only on the scale $\delta t$. 
From the Green--Kubo expression for the friction,
$2k_BT\gamma = \int_{-\infty}^\infty\!\langle f_{\mathrm{int}}(0)
f_{\mathrm{int}}(t)\rangle\,dt=A$, so
\begin{equation}
  \langle f_{\mathrm{int}}(t)f_{\mathrm{int}}(t')\rangle
  = 2k_BT\,\gamma\,\delta_\tau (t-t').
  \label{deltaFs}
\end{equation}
We point out that, unlike heavy-particle Brownian motion, there is no timescale separation between the decorrelations of the velocity and of the Langevin force.
\subsection{Collective variables}
We now consider a wire segment of length $\delta x$, macroscopically short but much longer than $\max (\ell ,\tau\sqrt{k_BT/m})$, and sufficiently long to ignore fluctuations in the number of particles in it, $n(x)\delta x$. 

From (\ref{defI}), omitting the index $x$, invoking molecular chaos ($\langle v_i(0)v_j(t)\rangle =\delta_{ij}\langle v_i(0)v_i(t)\rangle$, where $v_i$ is the velocity of particle $i$; this is expected because the velocities of different particles immediately after a collision are uncorrelated), and using (\ref{Ct}), we obtain
\begin{equation}
    \langle I(x,0)I(x,t)\rangle=\frac{q^2}{\delta x^2}\sum_{i,j}\langle v_i(0)v_j(t)\rangle =\frac{n(x)q^2}{\delta x}\frac{k_BT\tau}{m}\delta_\tau (t)\;.
\end{equation}
Here, the function $\delta_\tau$ is microscopically different but macroscopically equivalent to the function with the same name that appears in (\ref{deltaFs}). In view of (\ref{Drude}), this expression becomes
\begin{equation}
    \langle I(x,0)I(x,t)\rangle=\frac{2k_BT}{\rho (x)\delta x}\delta_\tau (t)\;,
    \label{I0It}
\end{equation}
where $\rho (x)\delta x$ is the resistance of the segment. In Rytov's theory \cite{Ry}, the current densities at points separated by a macroscopic distance are taken as uncorrelated, but this situation is beyond the scope considered here, which is limited to distances where Kirchhoff's law is partially obeyed.

Consider the system of $n(x)\delta x$ particles in the segment and study the equation of motion of its center of mass. The mass becomes $mn(x)\delta x$ and we denote the total Langevin force on all these particles by $F_{\mathrm{int}}$. The friction coefficient becomes $\Gamma =2mn(x)\delta x/\tau$ and (\ref{deltaFs}) becomes $\langle F_{\mathrm{int}}(0)F_{\mathrm{int}}(t)\rangle
  = [4k_BTmn(x)\delta x/\tau]\delta_\tau (t)$. Regarding $F_{\mathrm{int}}$ as an electric force exerted by a fluctuating electric field $E(x,t)$ (which is induced by the lattice and by the particles close to the segment, keeping track of the $x$-component only), $F_{\mathrm{int}}(t)=n(x)\delta x\,qE(x,t)$ and
  $\langle E(x,0)E(x,t)\rangle = [4k_BTm/q^2n(x)\delta x\tau]\delta_\tau (t)$, which in view of (\ref{Drude}) is $\langle E(x,0)E(x,t)\rangle =[2k_BT\rho (x)/\delta x]\delta_\tau (t)$. 
  If one furthermore assumes that the equilibrium Langevin fields acting on different coarse-graining cells are statistically independent, one obtains
    \begin{equation}
      \langle E(x,0)E(x',t)\rangle = 2k_BT\rho (x)\delta_{\delta x}(x-x')\delta_\tau (t)\;,
      \label{E0Et}
  \end{equation}
  where $\delta_{\delta x}$ stands for a macroscopic $\delta$-function with width $\delta x$.

  If instead of our modified Drude model we consider, e.g. the standard Drude model, the friction coefficient has a different value, but since $\rho\propto\gamma$, (\ref{I0It}) and (\ref{E0Et}) remain unchanged for any value of $\gamma$.



\begin{thebibliography}{9}
\bibitem{ein}A. Einstein, On the theory of Brownian motion, Ann. Physik \textbf{19} 371--381 (1906).
\bibitem{john} J. B. Johnson, Thermal agitation of electricity in conductors, Phys. Rev. \textbf{32} 97--109 (1928).
\bibitem{ny} H. Nyquist, Thermal agitation of electric charge in conductors, Phys. Rev. \textbf{32} 110--113 (1928).
\bibitem{Qu} J. F. Qu, S. P. Benz, H. Rogalla, W. L. Tew, D. R. White, and K. L. Zhou, Johnson noise thermometry, Meas. Sci. Technol. \textbf{30} 112001 (2019).

\bibitem{Urbina}F. C. Wellstood, C. Urbina, and J. Clarke, Hot-electron effects in metals, Phys. Rev. B \textbf{49}, 5942 (1994).
\bibitem{Dej} F. K. Dejene, J. Flipse, G. E. W. Bauer and B. J. van Wees, Spin heat accumulation and spin-dependent temperatures in nanopillar spin valves, Nature Physics \textbf{9}, 636-639 (2013).
\bibitem{Loss} E. V. Sukhorukov and D. Loss, Noise in multiterminal diffusive conductors: Universality, nonlocality and exchange effects, Phys. Rev. B \textbf{59} 13054 (1999).
\bibitem{Brian}J. Waissman et \textit{al}., Electronic thermal transport measurement in low-dimensional materials with graphene non-local noise thermometry, Nature Nanotechnology \textbf{17}, 166 (2022).
\bibitem{Nature} O. S. Lumbroso, L. Simine, A. Nitzan, D. Segal, and O. Tal, Electronic noise due to temperature difference demonstrated in molecular junctions: beyond standard thermal and shot noises, Nature \textbf{562}, 240 (2018).
\bibitem{Kubo}R. Kubo, The fluctuation-dissipation theorem, Rep. Prog. Phys. \textbf{29} 255–284 (1966).
\bibitem{Jan} J. V. Sengers, Mass and thermodiffusion in non‑equilibrium fluctuating hydrodynamics, Int. J. Thermophys. 45:132 (2024).
\bibitem{hydro}D. Bedeaux, S. Kjelstrup, and J.V. Sengers, eds., Non-equilibrium Thermodynamics with Applications (IUPAC, RSC Publishing, Cambridge, 2016).
\bibitem{BK} J. Berger and G. Katriel, Nonlocal origin and correlations in the Johnson noise at nonuniform temperature, Phys. Rev. B \textbf{112}, 224311 (2025).
\bibitem{Bellon} B. Monnet, S. Ciliberto and L. Bellon, Extended Nyquist formula for a resistance subject to a heat flow, J. Stat. Mech. (2019) 104011.
\bibitem{prl}S. Ciliberto, A. Imparato, A. Naert, and M. Tanase, Heat Flux and Entropy Produced by Thermal Fluctuations, Phys. Rev. Lett. \textbf{110}, 180601 (2013).
\bibitem{Berut} A. B\'{e}rut, A. Imparato, A. Petrosyan, and S. Ciliberto, The role of coupling on the statistical properties of the energy fluxes between stochastic systems at different temperatures, J. Stat. Mech. (2016) 054002.
\bibitem{liquid} J. A. Barker  and D. Henderson, What is ``liquid"? Understanding the states of matter, Rev. Mod. Phys. \textbf{48}, 587 (1976).
\bibitem{Giamarchi} T. Giamarchi, Quantum Physics in One Dimension (Clarendon, Oxford, 2003).
\bibitem{Clarke} R. F. Voss and J. Clarke, Flicker (1/f) noise: Equilibrium temperature and resistance fluctuations, Phys. Rev. B \textbf{13}, 556 (1976).
\bibitem{Maz} S.R. de Groot and P. Mazur, Non-Equilibrium Thermodynamics (North Holland, Amsterdam, 1962, Reprinted by Dover, 1984)
\bibitem{Lk} R. Meservey and P. M. Tedrow, Measurements of the Kinetic Inductance of Superconducting Linear Structures, J. Appl. Phys. \textbf{40}, 2028 (1969).
\bibitem{Ry} S. M. Rytov, Y. A. Kravtsov, and V. I. Tatarskii, Principles of Statistical Radiophysics, Vol. 3 (Springer-Verlag, 1989).
\bibitem{Henkel} C. Henkel, Nano-scale thermal transfer – an invitation to fluctuation electrodynamics, Z. Naturforsch. A \textbf{72}, 99 (2017).
\bibitem{Abdalla}S.-A. Biehs and P. Ben-Abdallah, Fluctuations of radiative heat exchange between two bodies. Phys. Rev. B \textbf{97}, 201406 (2018).

\end{thebibliography}
\end{document}